# Observation of Coherence in the Photosystem II Reaction Center

Franklin D. Fuller, Jie Pan, S. Seckin Senlik, Daniel E. Wilcox and Jennifer P. Ogilvie*

*Department of Physics and Biophysics, University of Michigan, Ann Arbor, MI 48109*

**Photosynthesis powers life on our planet. The basic photosynthetic architecture comprises antenna complexes to harvest solar energy and reaction centers to convert the energy into a stable charge separated state. In oxygenic photosynthesis, the initial charge separation event occurs in the photosystem II reaction center; the only known natural enzyme that uses solar energy to split water. Energy transfer and charge separation in photosynthesis are rapid and have high quantum efficiencies. Recently, nonlinear spectroscopic experiments have suggested that electronic coherence may play a role in energy transfer efficiency in antenna complexes. Here we report the observation of coherence in the photosystem II reaction center by two dimensional electronic spectroscopy. The frequencies of the observed coherences match exciton difference frequencies and/or known vibrational modes of the photosystem II reaction center. These observations raise questions about the possible role of electronic and/or vibrational coherence in the fundamental charge separation process in oxygenic photosynthesis.**

Photosynthesis is the process by which plants and other photosynthetic organisms convert solar energy into chemical energy. Although the photosynthetic machinery varies among organisms, the basic architecture consists of light-harvesting antennae arrays that gather solar energy and funnel it to reaction centers (1). Within reaction centers the energy is converted to a charge separated state that drives the later stages of photosynthesis. In oxygenic photosynthesis in plants, algae and cyanobacteria the first charge separation event occurs in the Photosystem II reaction center (PSII RC). The PSII RC is unique among biological systems in its ability to use solar energy to split water, making its function of particular interest for designing artificial photosynthetic devices (2).

The relatively recent development of two-dimensional electronic spectroscopy (2DES) has provided an incisive tool for studying the energy transfer pathways and probing the electronic structure of photosynthetic complexes (3-7). Compared to pump-probe spectroscopy, 2DES adds an extra dimension: the excitation frequency axis, which enables correlations to be made between excitation and detection frequencies. This added dimension provides a more direct view of populations and coherences that appear on the diagonal and off-diagonal regions of a 2D spectrum. The first application of 2DES to the study of the Fenna-Matthews-Olson complex (FMO) revealed its utility in picking apart complex energy transfer pathways (5). Subsequent work on FMO revealed coherent oscillations in the 2DES data as a function of population time, and these oscillations were assigned to electronic coherences (8). Coherences in 2DES data have now been observed in FMO at room temperature (9), and have been seen in a number of other photosynthetic systems (10-13), as well as J aggregates (14) and polymers (15). The possibility of long-lived electronic coherence has captured the imagination of the theoretical community, inspiring numerous works to explain the origin of the 2DES observations and to determine their possible relevance to efficient energy transfer (16-20), and more recently electron transfer (21).

While the bulk of 2DES experiments have focused on light harvesting antenna, the 2DES method has recently been applied to reaction centers. In 2010 we reported the first 2DES reaction center studies (22), examining the energy transfer and charge separation processes in the D1D2 cytb559 PSII RC (structure shown in Figure 1). A recent report used 2DES to study the bacterial reaction center (BRC) in which the primary electron donor was oxidized, blocking the charge separation (23). An earlier two color photon echo paper also studied the oxidized BRC (24), and both of these studies reported electronic coherences. Pump-probe experiments on the BRC have reported coherences that were considered to be vibrational (25-31). Analogous pump-probe experiments have not to date reported coherences in the PSII RC (32; 33), although they have been observed recently via 2DES (34; 35). Our initial PSII work (22) and subsequent modeling has focused on testing existing excitonic models of the PSII RC (36), and extending these models using a tight-binding formalism to describe the charge separation process (37). Our initial studies using 25 fs pulses provided inconclusive data regarding the presence or absence of coherences (22). Here, using a modified, higher signal-to-noise 2DES setup with 12 fs pulses (38), we report clear coherent signatures in the 2DES spectra of the PSII RC. We also discuss the possible physical origins of the coherences and speculate about their relevance to the essential charge separation function in the PSII RC.

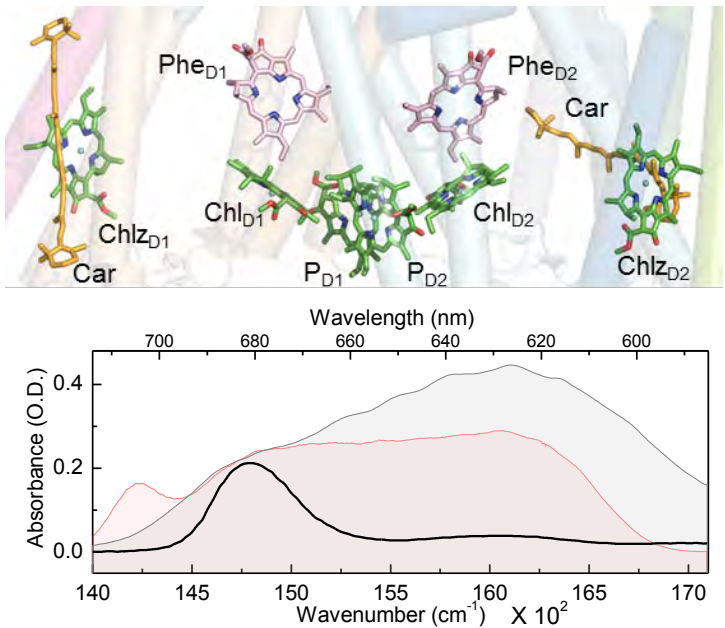

**Figure 1.** Top: Structure of the D1D2 cytb559 PSII RC (3ARC(39)), showing chlorophyll (Chl) molecules in green, pheophytin (Phe) molecules in light magenta, and carotenoid (Car) in orange. For clarity, phytyl tails of all chlorins were truncated. Bottom: Linear absorption spectrum of PSII RC at room temperature (black curve), and the employed excitation (red filled curve) and probe (grey filled curve) laser spectra.

**Results**

Figure 2A shows a typical absorptive 2DES spectrum of the PSII RC at a population time of 170 fs. Marked on the spectrum (arrows) are several diagonal and off-diagonal locations where we observe strong coherences as a function of the population time ($t_2$) as shown in Figure 2B. The frequency content of these individual time traces is shown in Figure 2C, where the Fourier transform has been taken with respect to $t_2$. Figure 3D shows a "summary" spectrum obtained upon summing (over $\omega_1$ and $\omega_3$) the square of the Fourier transforms of the $t_2$ traces at each ($\omega_1$, $\omega_3$) value in the 2D spectrum, yielding a single spectrum along $\omega_2$. Thus this spectrum displays the dominant frequencies of the coherences observed throughout the entire 2D spectrum. Above the summary spectrum are bar graphs that illustrate the exciton difference frequencies predicted by the Novoderezhkin (40), and modified Novoderezhkin (36) models. Also shown are known vibrational frequencies of the PSII RC from fluorescence line-narrowing experiments (41) and surface-enhanced resonance Raman (42) studies of the PSII RC. To reveal the distribution of the observed frequencies throughout the 2D spectrum we show "coherence amplitude maps" in Figure 4 for the dominant modes indicated in Figure 3D. The coherence amplitude maps are slices (at frequency $\omega_2$) through the three dimensional solid obtained upon Fourier transforming the time-ordered array of 2DES data with respect to $t_2$. Further details of how the maps are obtained are given in the Methods section. Also shown on each coherence map are dotted lines parallel to the diagonal, offset from it by the appropriate coherence frequency. These lines are shown to aid in discerning the physical origin of the coherences according to different protocols in the literature (11; 43-45).

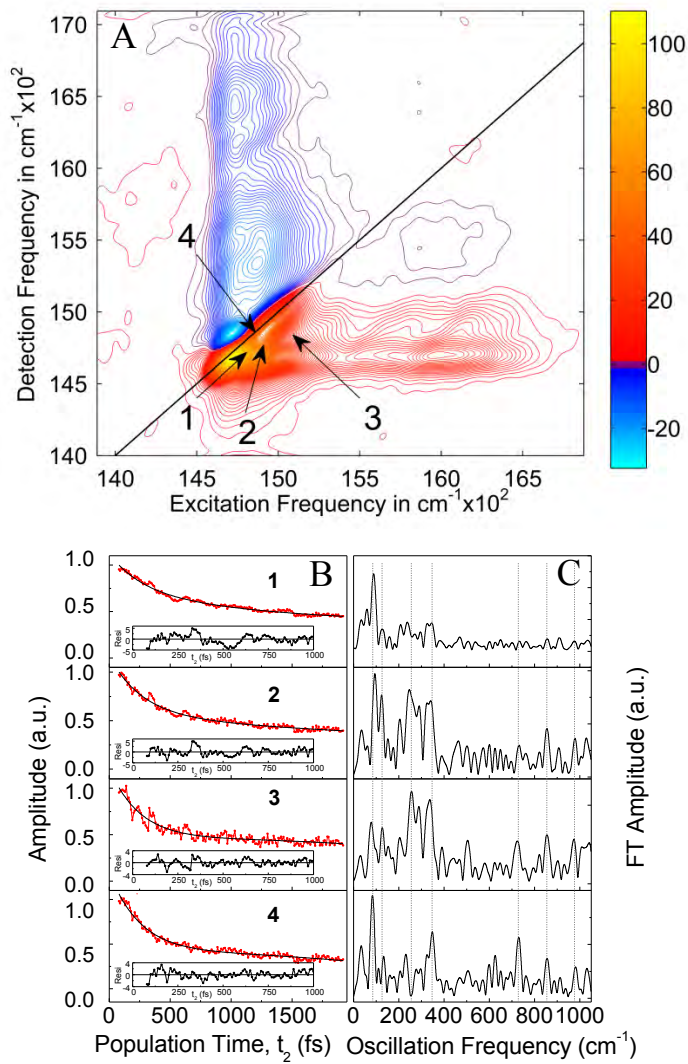

**Figure 2.** A: Absorptive (real, phased) 2D spectrum of the PSII RC at 77K at a population time of 170 fs. B: Representative population time ($t_2$) traces taken at the locations indicated in A by the numbered arrows. Data is shown in red, with background fit in black. The insets show the data after subtraction of the background. C: The corresponding Fourier transforms of the population time traces in B (taken after background subtraction). Gray vertical lines show locations of peaks found consistently throughout the 2D spectrum.

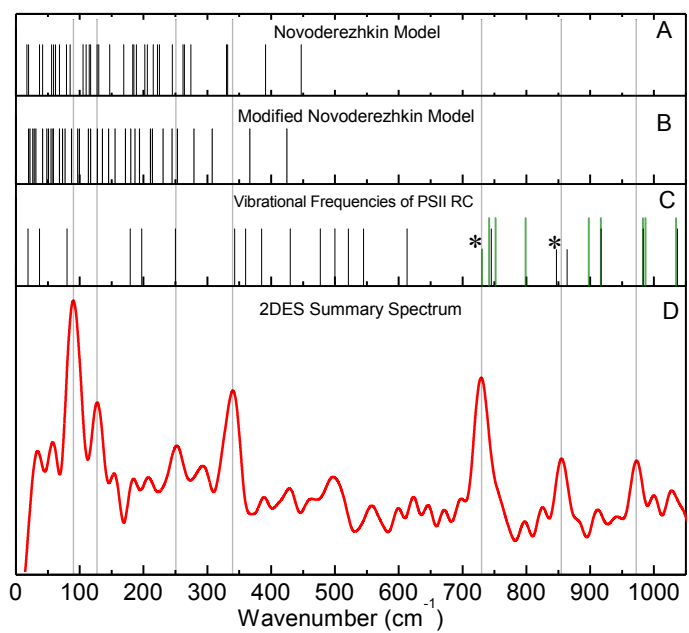

**Figure 3:** Exciton difference frequencies for the A: Novoderezhkin (40) and B: modified Novoderezhkin(36) models of the PSII RC. C: vibrational frequencies for the PSII RC from fluorescence line-narrowing(41) (black) and surface-enhanced resonance Raman (42) (green) experiments. Note: * indicates low amplitudes of transitions as compared to other peaks. D: Summary spectrum as a function of $\omega_2$, produced by summing (over $\omega_1$ and $\omega_3$) the square of the amplitude of the 2D spectrum at each $\omega_2$ point. Gray lines indicate the positions of major peaks (91 cm$^{-1}$, 127 cm$^{-1}$, 251 cm$^{-1}$, 339 cm$^{-1}$, 730 cm$^{-1}$, 854 cm$^{-1}$, 974 cm$^{-1}$).

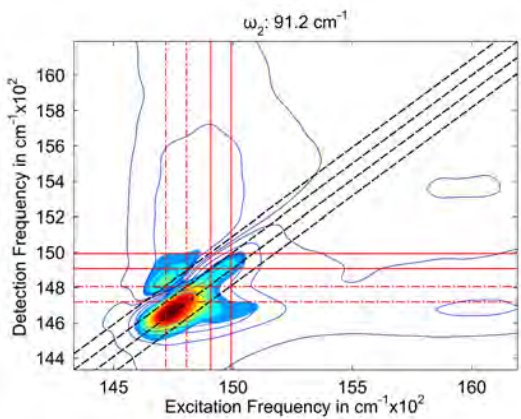
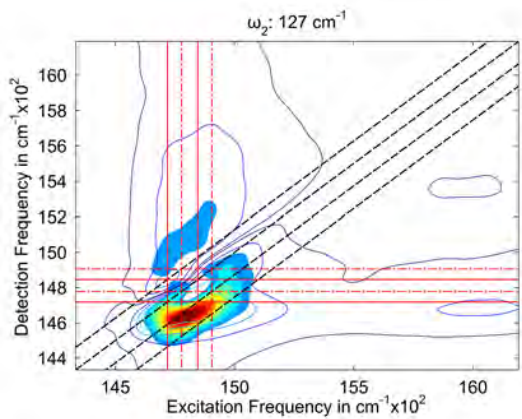
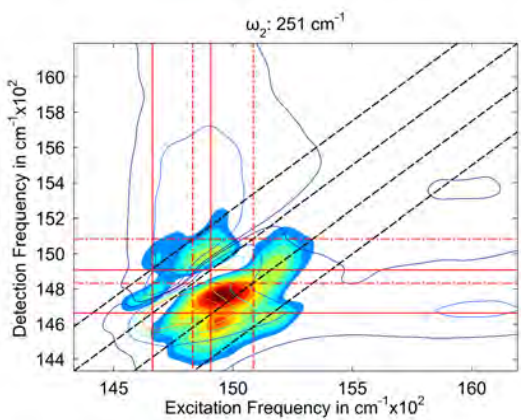
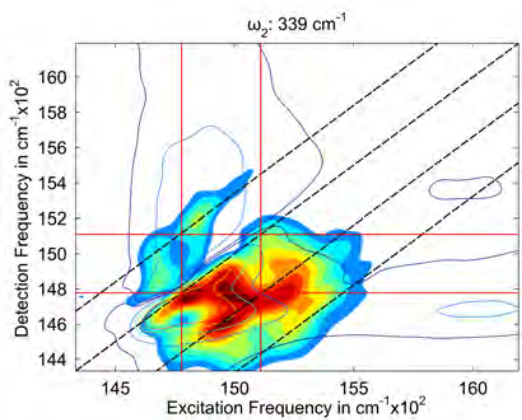
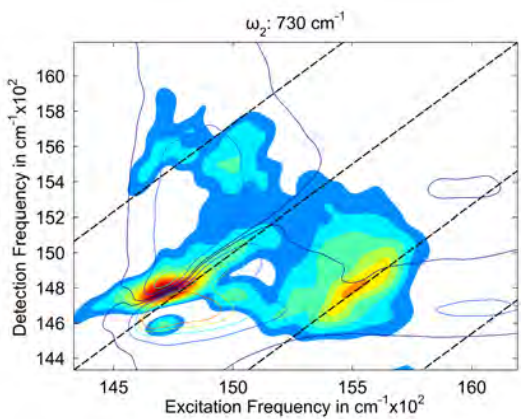
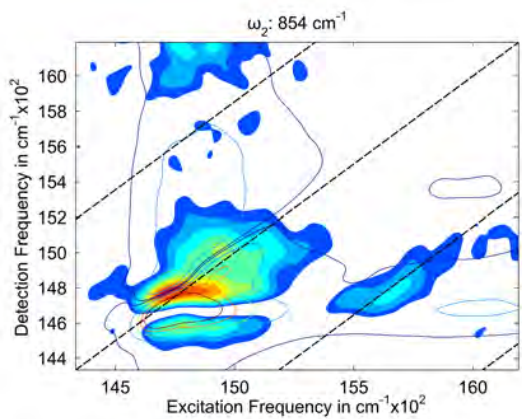
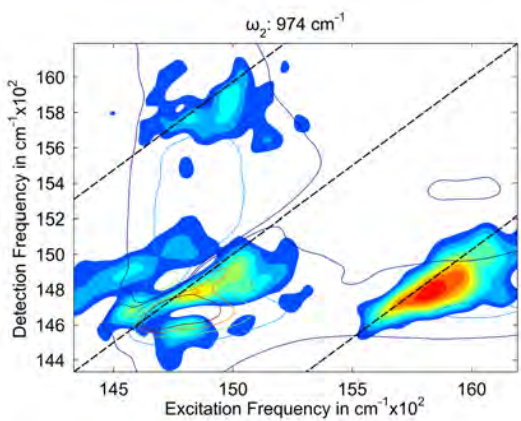

**Figure 4:** Coherence amplitude maps (filled contours) derived from the absorptive 2D spectra by taking slices at particular "population frequency" $\omega_2$ values, indicating the dependence of the observed coherences on the excitation and detection wavelength. Coherence amplitude maps at 91 cm$^{-1}$, 127 cm$^{-1}$, 251 cm$^{-1}$, 339 cm$^{-1}$, 730 cm$^{-1}$, 854 cm$^{-1}$ and 974 cm$^{-1}$. For each map dashed black lines indicate the diagonal and parallel lines offset from the diagonal by $\pm\omega_2$ and $-2\omega_2$. Red lines indicate the location of exciton pairs with difference frequency matching the $\omega_2$ frequency (solid red – Novoderezhkin, dashed red – modified Novoderezhkin model). Where multiple exciton pairs had matching difference frequencies (within 10 cm$^{-1}$) the closest match is shown. Overlaid open contours show the absorptive 2D spectrum, averaged over population time $t_2$.

**Discussion**

As shown in Figures 2-4, the 2DES data reveal clear coherences in the PSII RC at a number of frequencies. Here we first briefly review the current understanding of charge separation in the PSII RC. We then speculate about the importance of electronic and vibrational coherence to the charge separation process. Finally, we discuss the experimental data and physical origins of the observed coherences, considering the possible functional relevance of the coherences within the context of different exciton models of the PSII RC.

**Relevance to Charge Separation:**

It has been previously proposed that the initial charge separation in the PSII RC occurs between Chl$_{D1}$ and Phe$_{D1}$ (46). This proposal has been supported by mutant studies(47) and other work (48-51). Others argue that at room temperature this pathway is not active and the dominant mechanism involves charge separation between the excited P$_{D1}$P$_{D2}$ pigments and Phe$_{D1}$ (52) with possible involvement of Chl$_{D1}$ as the primary electron acceptor (32; 33). It has also been proposed that both of these two charge separation pathways are active (53-55); a "Chl$_{D1}$" pathway: (Chl$_{D1}$Phe$_{D1}$)* → Chl$_{D1}{}^+$Phe$_{D1}{}^-$ → P$_{D1}{}^+$Phe$_{D1}{}^-$ as well as a "$P_{D1}$" pathway: $(P_{D1}P_{D2}Chl_{D1})^* \rightarrow P_{D2}^+P_{D1}^- \rightarrow P_{D1}^+Chl_{D1}^- \rightarrow P_{D1}^+Phe_{D1}^-$ where * denotes the initially excited pigments that precede the charge separation. In both the PSII RC and the BRC the timescale for primary charge separation is thought to be several picoseconds, coincident with the observed coherences (32; 33; 48; 53; 55-57).

Electronic coherence, defined here as coherence between pure exciton states without explicit coupling to vibrational states, could be functionally relevant if it promotes effective transfer of excitation to the appropriate pigments to set the stage for charge separation. This interpretation has been suggested in the recent study of the oxidized bacterial reaction center (23). If multiple pathways to charge separation exist, then quantum interference between the pathways could result in coherences. Pathway interferences have been recently studied in simulations of electron transfer in the photosystem I reaction center, where they have been shown to be able to enhance electron transfer efficiency (21). Whether or not multiple pathways to charge separation are active within single PSII RCs at room temperature is unknown.

Vibrational coherences could also potentially influence charge separation in reaction centers. The long-lived low frequency coherences in the BRC observed in 1991 by Vos *et al.* (25-28) and subsequently other groups (29-31) using pump-probe spectroscopy were attributed to vibrational coherence (28). Theoretical work has investigated the role of vibrational and/or electronic coherence in electron transfer (58-60). Applications to the BRC have reached differing conclusions regarding the possible importance of vibrational coherence to electron transfer efficiency (61-63). Extensive studies of BRC mutants demonstrate that protein dynamics dictates the rate of electron transfer (64), but the nature of these dynamics is not yet understood. We note that low frequency modes have been implicated in electron transfer in a completely different system, cytochrome P450, based on pump-probe experiments (65). It is quite likely that, while the detailed mechanisms may be different, the underlying role of coherence may be similar in these systems and the PSII RC.

**Physical Origin:**

Electronic coherence between excitonic states should manifest itself as modulation in the 2D spectrum as a function of $t_{2(8)}$. This modulation should occur at a frequency that corresponds to the difference frequency between the participating excitonic states. We find that several of the dominant frequencies present in the 2D spectrum, shown in Figure 3D, match exciton difference frequencies predicted by the Novoderezhkin and modified Novoderezhkin excitonic models. Where close matches occur (within 10 cm$^{-1}$) we indicate the exciton lines on the coherence amplitude maps in Figure 4. Also shown in Figure 4 are dashed lines immediately above and below the diagonal, offset by the $\omega_2$ frequency, indicating off-diagonal positions where potential electronic coherences could appear. Details of the exciton models are given in Tables S1 and S2 in the Supplementary Materials, where we show the pigments involved in each exciton state, the exciton energies and difference frequencies, noting which ones we observe in our experiment. Figure 3 conveys a subset of this information in a condensed format. Briefly, the Novoderezhkin (40; 66) model was derived using an evolutionary algorithm to find model parameters that would simultaneously fit multiple linear spectroscopy measurements, Stark spectroscopy and pump-probe data. The modified Novoderezhkin

(36) model adjusted the original parameters of the Novoderezhkin model to obtain a better match between simulated 2DES spectra and experimental data. These changes included altering the identity of the charge transfer state, lowering the degree of disorder, and increasing the system-bath coupling. In addition, a higher-resolution crystal structure (39) was used for calculating the coupling between pigments. It can be seen from Figure 3 that the four lowest frequency modes (91 cm$^{-1}$, 127 cm$^{-1}$, 251 cm$^{-1}$, 339 cm$^{-1}$) match exciton difference frequencies within both models. We note that we also examined a third excitonic model by Raszewski *et al.* (51), finding a single match to the 339 cm$^{-1}$ mode).

In addition to matching excitonic difference frequencies, there have been several protocols developed to identify the origin of an observed coherence as electronic or vibrational. These have been based on simple models that separately consider vibrational and/or excitonic transitions (11; 43-45). A hallmark signature of purely electronic coherence is the absence of oscillation amplitude along the diagonal in the rephasing spectrum and a purely diagonal oscillation in the non-rephasing spectrum. In addition, vibrational coherence on the ground or excited electronic state is thought to give rise to signal below the diagonal in the rephasing spectrum (located along the lowest dashed line parallel to the diagonal) (11; 44). It can be seen in Figure S1 in the Supplementary Materials, that our data definitively supports a vibrational origin for higher frequency modes (above and including 339 cm$^{-1}$), as a strong diagonal amplitude is seen in the rephasing coherence amplitude maps. In the 339 cm$^{-1}$ map the rephasing spectrum shows amplitude along the lowest dashed line, also supporting a vibrational assignment. The bandwidth limits of our detection axis preclude similar tests of the higher frequency modes. The modes above and including 339 cm$^{-1}$ correspond reasonably well with ground state vibrational modes observed in fluorescence line-narrowing (41) and surface enhanced resonance Raman spectroscopy (42) of the PSII RC as indicated in Figure 3. Deviations from the known modes may arise from excited state coherences that are shifted with respect to ground state vibrations. We note that recent 2DES studies of J-aggregates have detected coherence at a frequency that was absent in nonresonant Raman studies (14).

For the lower frequency modes, the picture is less clear. At 127 cm$^{-1}$, we see a suggestive lack of diagonal amplitude in the rephasing and little cross peak amplitude in the non-rephasing spectra. This, combined with the fact that no vibrational lines from fluorescence line-narrowing overlap the 127 cm$^{-1}$ mode lend credence the assignment of this as an electronic coherence. We note that the surface-enhanced resonance Raman study of the PSII RC is limited to frequencies above 500 cm$^{-1}$ and therefore does not rule out a 127 cm$^{-1}$ vibrational mode. The other low frequency modes (91 cm$^{-1}$ and 251 cm$^{-1}$) show considerable amplitude near the diagonal in the rephasing spectrum and the fluorescence line-narrowing data shows vibrational modes in close proximity. Within the modified Novoderezhkin and Novoderezhkin exciton models, the 91 cm$^{-1}$ and 251 cm$^{-1}$ coherences involve one exciton in which the peripheral chlorophyll (Chlz$_{D1}$, Chlz$_{D2}$) and/or pigments on the D2 side are excited, while the second exciton involves either the CT state and/or the pigments involved in P$_{D1}$ or Ch$_{D1}$ pathways. These coherences could be interpreted as enabling effective transfer of energy to the pigments in the P$_{D1}$ and Ch$_{D1}$ pathways to set the stage for charge separation. This is depicted in Figure 5A, where we show the pigments involved in the 251 cm$^{-1}$ coherence between excitons 5 and 9 within the modified Novoderezhkin model.

Within both exciton models there are two difference frequencies that closely match the 127 cm$^{-1}$ mode. In both models one of these exciton pairs (excitons 2 and 6 in the modified, 3 and 6 in the original Novoderezhkin model) involves coherence between pigments and charge transfer states involved in the Chl$_{D1}$ and P$_{D1}$ charge separation pathways. We depict the pigments involved in this excitonic coherence within the modified Novoderezhkin model in Figure 5B. Within this model, exciton 2 involves excitation of Chl$_{D1}$ and Phe$_{D1}$, the excited pigments that initiate charge separation via the Ch$_{D1}$ pathway. In contrast, exciton 6 involves excitation of P$_{D1}$ and the charge transfer state, suggesting its importance for initiating charge separation via the P$_{D1}$ pathway. The fact that the 127 cm$^{-1}$ coherence involves both Chl$_{D1}$ and P$_{D1}$ pathways raises the possibility of quantum mechanical interference influencing charge separation efficiency.

The 339 cm$^{-1}$ mode corresponds well to two exciton difference frequencies in the Novoderezhkin model. One of these involves a coherence between a peripheral chlorophyll and the charge transfer state. Like the 127 cm$^{-1}$, the other coherence is between two excitons that involve excitations that initiate charge separation via the Chl$_{D1}$ and P$_{D1}$ pathways.

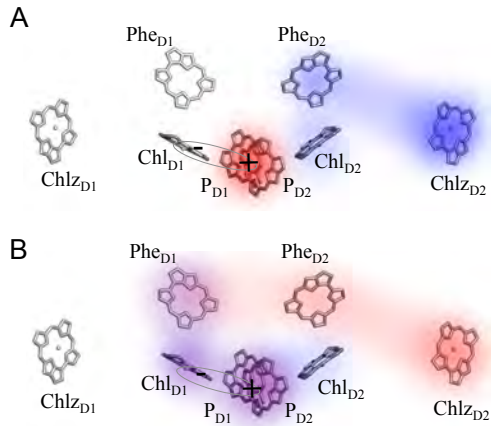

**Figure 5:** A: Depiction of the excitonic coherence of the 251 cm$^{-1}$ mode within the modified Novoderezhkin model, showing the pigments participating in excitons 5 (blue) and 9 (red). B: Depiction of the excitonic coherence of the 127 cm$^{-1}$ mode within the modified Novoderezhkin model, showing the pigments participating in excitons 2 (blue) and 6 (red). Pigments are colored according to their degree of participation in the two excitonic states, with darker coloring indicating higher participation ratio. The charge transfer state $(P_{D1}P_{D2})^+Chl_{D1}^-$ is indicated by the gray oval and charges on the appropriate pigments.

**Mixed electronic and vibrational (vibronic) contributions:**

Like the coherences that have been reported in other photosynthetic systems, the coherences we observe here persist on picosecond timescales at 77 K. There has been considerable theoretical work towards understanding the physical origin of long-lived electronic coherence in light-harvesting systems. Some of this work has predicted electronic coherence lifetimes of several hundreds of femtoseconds (24; 67; 68), while models that propose correlations between site energy fluctuations of different pigments have succeeded in reproducing the ~picosecond lifetime of the observed coherences (17; 24; 67; 69-72). Molecular dynamics simulations of FMO have not supported the correlated site energy hypothesis (73; 74).

Recent theoretical and experimental work has questioned the idea that the coherences observed in 2DES experiments on light-harvesting systems can be classified as purely electronic or vibrational (14; 75-78). Coupling between excitons and vibrations has been invoked as a mechanism for fast and effective energy distribution in cryptophyte algae (79; 80). Tiwari et al. (81) suggest that such resonances can help drive nonadiabatic energy transfer. Their work, in which the observed coherences are interpreted as arising from ground-state vibrational wavepacket motion, has successfully reproduced key signatures in the experimental data on the FMO complex. In our data, the 91 $cm^{-1}$ and 251 $cm^{-1}$ coherences in particular likely have mixed vibrational and electronic (vibronic) origins.

**Conclusions:**

We have observed coherent dynamics in the PSII RC via 2DES. These coherences are found throughout the 2D spectrum, at a number of frequencies. Some of these correspond well with exciton difference frequencies and in some cases the excitons involve participation from the pigments that are thought to be key to charge separation in the PSII RC. Most of the observed coherences match known vibrational modes. A growing body of work has begun to explore the role of vibrational degrees of freedom in explaining the long-lived coherences seen in FMO and other light-harvesting systems. In current excitonic models of the PSII RC, vibrational degrees of freedom typically enter via the spectral density and are not explicitly included in the excitonic Hamiltonian. Recently Abramavicius et al. considered the possibility of observing electronic coherence in the PSII RC (82), finding long-lived coherences at diagonal and off-diagonal locations. Their work employed a simplified exciton model that did not include vibrational states explicitly, either in the system Hamiltonian or the spectral density. They also did not consider the relevance of coherence to the efficiency of charge separation, an idea that has been explored in simulations of the photosystem I reaction center (21). The debate continues regarding the significance of electronic and/or vibrational coherence to energy transfer in photosynthesis. This debate is relevant to reaction centers because the energy transfer process sets the stage for efficient charge separation. Given the similarity between the frequencies of the coherences we observe with known vibrational modes, it is possible that vibrational-excitonic resonance is important for effective energy transfer in the PSII RC. Beyond effectively channeling the excitation energy to the appropriate pigments for charge separation, it is also possible that the coherences we observe are significant for the charge separation mechanism itself. Key to resolving these issues are further theoretical and experimental work to determine the signatures of electronic, vibrational and vibronic coherences in 2DES spectra. Careful 2DES studies of the monomers and reaction center mutants should provide further insight into the possible significance of coherence to the energy transfer and charge separation processes of the PSII RC.

## Acknowledgments

F. D. Fuller, J. Pan and J. P. Ogilvie performed design and execution of the experiments and interpretation of the data. They gratefully acknowledge the support of the Office of Basic Energy Sciences, U.S. Department of Energy (grant #DE-FG02-07ER15904). S. S. Senlik, who aided in data collection and debugging of the optical setup acknowledges the National Science Foundation (grant # PHY-0748470). D. E. Wilcox, who contributed to data fitting and debugging of the optical setup acknowledges the support of the Center for Solar and Thermal Energy Conversion (CSTEC), an Energy Frontier Research Center funded by the U.S. Department of Energy (DOE), Office of Science, Basic Energy Sciences (BES), under award DE-SC0000957.

**Supplementary Materials**

**Methods**

Tris-washed BBY particles were extracted from commercially available spinach (83), following which the D1-D2-cyt *b*559 reaction centers were isolated using the approach of van Leeuwen *et al.* (84). Prior to use the samples were diluted with a sucrose-free Bis-Tris solution and concentrated with a spin filter (Millipore). Glycerol was added to produce a glycerol:buffer ratio of 2:1 (v/v), and the sample was vacuum degassed prior to being sealed in an optical cell with a sample thickness of 380 μm and an OD of ~0.2 at 680 nm.

The 2DES measurements were made using a hybrid diffractive optic (85) and pulse-shaping based (86; 87) approach that combines the advantages of background-free detection with the precise time-delays and phase-cycling capabilities of a pulse-shaper (38). We briefly describe the setup here. The laser source consists of a Ti:Sapph oscillator (MaiTai SP) seeding a regenerative amplifier (Spectra Physics Spitfire Pro). The 4 mJ, 500 Hz, 800 nm output is split and feeds two non-collinear optical parametric amplifiers (NOPAs) tuned to 680 nm (Figure 1B shows the pump and probe spectra used in the experiment). One beam, referred to here as the pump, is sent through a pre-compensating grism and then into acousto-optic pulse shaper (Dazzler, Fastlite) where it is compressed and split into two pulses with a programmable inter-pulse delay and phase. The second NOPA is compressed using a separate grism pair and is delayed by $t_2$ with respect to the pump pulses using a conventional delay stage. The pump and probe beams are directed into a diffractive optic imaging system where the crossing angle at the sample is approximately 1 degree. The pulse duration measured at the sample were 12 and 15 fs for pump and probe respectively. The radiant exposure of each pump pulse was 0.55 J/m$^2$ and 0.68 J/m$^2$ for the probe pulse, which corresponds to a 4% excitation probability per reaction center per pulse for both pump and probe, low enough to avoid exciton annihilation effects(88). The elimination of scatter from the 2DES data was achieved by a combination of a six phase-cycling scheme with the addition of chopping the probe pulse. For each 2D spectrum t1 was scanned to a maximum delay of 300 fs in increments of 1.85 fs. The coherence delay t1 was phase locked at 592.4 nm, such that the shortest period of any $\omega_1$ frequency within the pump bandwidth was 12.9 fs, yielding >3.4x Nyquist sampling of the excitation axis. The population time delay was scanned in 10 fs steps to a maximum delay of 1920 ps. The first 80 fs were not analyzed to avoid pulse-overlap effects. The frequency resolution of the reported Fourier transform spectra is 18 cm$^{-1}$. Figure 4 shows the pump and probe spectra used in the experiments, as well as an absorption spectrum of the PSII RC sample.

The 2DES spectra were phased to pump probe data using the projection slice theorem (89). 2DES spectra were also compared with previously published data acquired in the pump-robe geometry (22; 36) to confirm accurate phasing of the data.

**Coherence amplitude maps:** In order to resolve low frequency modes, population kinetics were removed using a global exponential fit (90). It was found that a single variable life time (266 fs) and two long fixed life times (2 ps and 10 ps) gave an adequate description of the population kinetics for the entire 2D spectrum over the ~2 picosecond scan range of the experiment. The population time trace for each frequency-frequency point in the 2D spectrum, after exponential kinetics were removed, is then Fourier-transformed along population time, and the amplitude of the Fourier transform is plotted at the frequency of interest. Fourier spectra were zero-padded to 1024 units after a one-sided Tukey window was applied in the time domain.

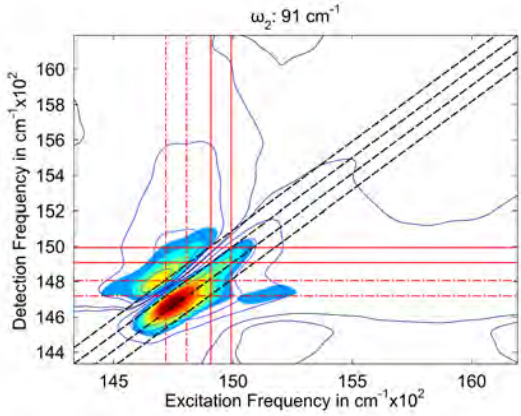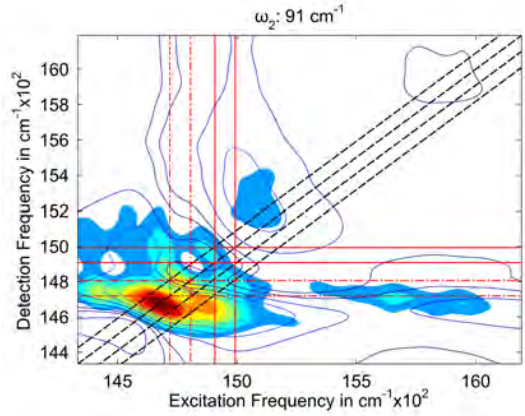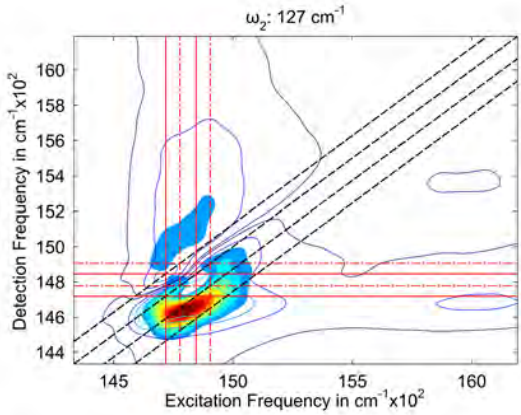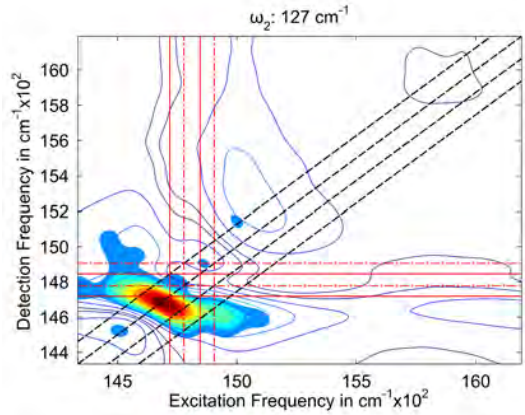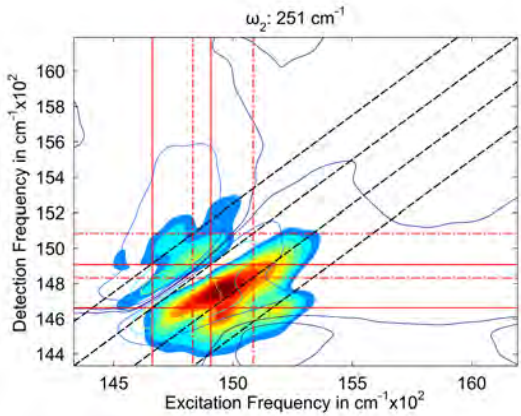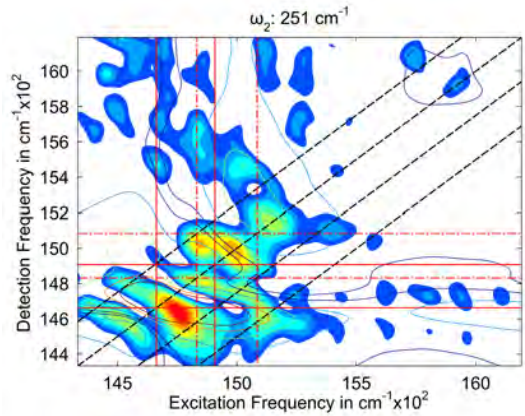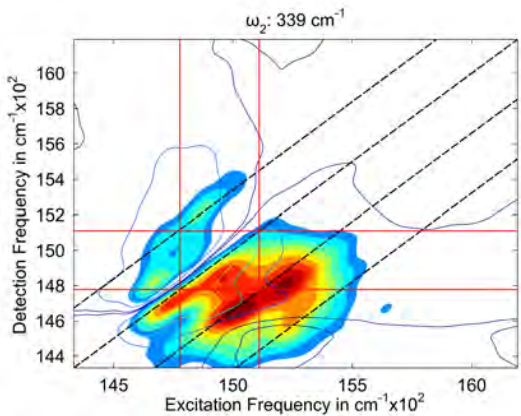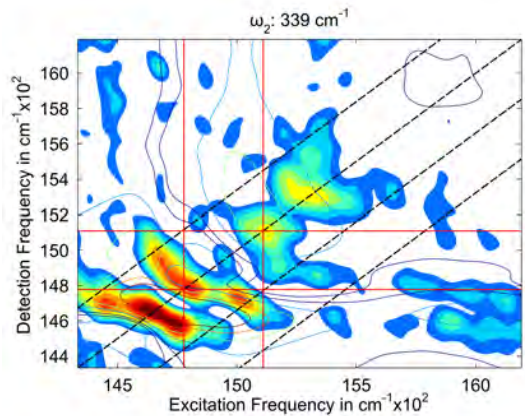

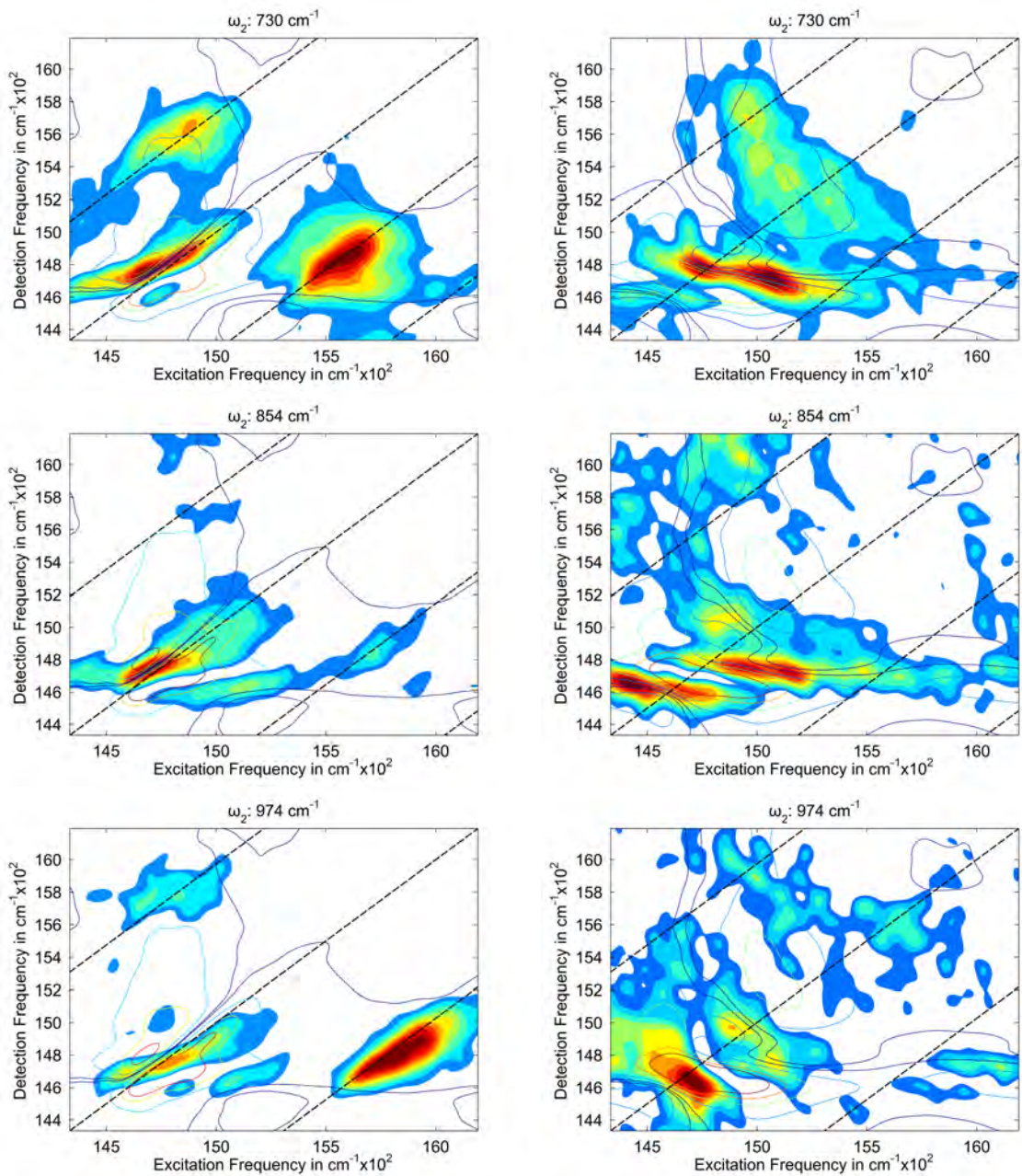

**Figure S1:** Real rephasing (left) and real nonrephasing (right) coherence amplitude maps (filled contours) derived from the 2D spectra by taking slices at particular "population frequency" $\omega_2$ values, indicating the dependence of the observed coherences on the excitation and detection wavelength. For each map dashed black lines indicate the diagonal and parallel lines offset from the diagonal by $\pm\omega_2$ and $-2\omega_2$. Red lines indicate the location of exciton pairs with difference frequency matching the $\omega_2$ frequency (solid red – Novoderezhkin, dashed red – modified Novoderezhkin model). Where multiple exciton pairs had matching difference frequencies (within 10 cm$^{-1}$) the closest match is shown. Overlaid open contours show the absorptive 2D spectrum, averaged over population time $t_2$.

**Exciton Models**

| Exciton No. | 1 | 2 | 3 | 4 | 5 | 6 | 7 | 8 | 9 |
|---|---|---|---|---|---|---|---|---|---|
| Pigments involved in excitonic state (participation ratios) | Chl $_{D1}$ (0.41) (P$_{D1}$P$_{D2}$)$^+$Chl$_{D1}^-$ (0.23) Phe$_{D1}$ (0.19) | (P$_{D1}$P$_{D2}$)$^+$Chl$_{D1}^-$ (0.28) Phe $_{D1}$ (0.23) Chl $_{D1}$ (0.22) Chl $_{D2}$ (0.10) | Chl$_{D2}$ (0.32) Phe$_{D2}$ (0.19) Phe$_{D1}$ (0.13) Chlz$_{D2}$ (0.11) | Chlz $_{D2}$ (0.24) Chl $_{D2}$ (0.22) Phe $_{D2}$ (0.21) | Chlz$_{D2}$ (0.34) Phe $_{D2}$ (0.16) Chl $_{D2}$ (0.11) | Chlz $_{D2}$ (0.23) Phe $_{D2}$ (0.15) Phe $_{D1}$ (0.12) P $_{D1}$ (0.11) (P$_{D1}$P$_{D2}$)$^+$Chl$_{D1}^-$ (0.10) | P $_{D1}$ (0.17) Phe $_{D2}$ (0.16) Chlz $_{D1}$ (0.14) Phe $_{D1}$ (0.11) P $_{D2}$ (0.10) Chl$_{D1}$ (0.10) | Chlz$_{D1}$ (0.81) | P $_{D1}$ (0.49) P $_{D2}$ (0.44) |
| 1 | 14661 | 58 | 117 | 145 | 172 | 194 | 214 | **245** | 425 |
| 2 |  | 14719 | 59 | **87** | 114 | **136** | 155 | 186 | 366 |
| 3 |  |  | 14778 | 28 | 55 | 77 | 97 | **128** | 308 |
| 4 |  |  |  | 14806 | 26 | 48 | 68 | 99 | 279 |
| 5 |  |  |  |  | 14832 | 22 | 42 | 73 | **253** |
| 6 |  |  |  |  |  | 14854 | 20 | 51 | 231 |
| 7 |  |  |  |  |  |  | 14874 | 31 | 211 |
| 8 |  |  |  |  |  |  |  | 14905 | 180 |
| 9 |  |  |  |  |  |  |  |  | 15085 |

**Table S1:** Exciton energies and difference frequencies (in cm$^{-1}$) in the Modified Novoderezhkin Model (36). Participation ratios are also shown for pigments involved in each exciton (only pigments with ≥10% participation ratio are shown). Difference frequencies matching the observed coherence frequencies (within 10 cm$^{-1}$) are marked in red.

| Exciton No. | 1 | 2 | 3 | 4 | 5 | 6 | 7 | 8 | 9 |
|---|---|---|---|---|---|---|---|---|---|
| Participating Pigments/CT states | $P_{D1}^+P_{D2}^-$ | $Chl_{D1}$ $Phe_{D1}$ $Phe_{D2}$ $P_{D1}$ $P_{D2}$ | $Chl_{D1}$ $Phe_{D1}$ $Phe_{D2}$ $Chl_{D2}$ | $Chl_{D1}$ $Phe_{D1}$ $Phe_{D2}$ $Chl_{D2}$ | $Chl_{D2}$ $P_{D1}$ $P_{D2}$ | $P_{D1}$ $P_{D2}$ $Chlz_{D2}$ | $P_{D1}$ $P_{D2}$ $Chlz_{D2}$ $Chlz_{D1}$ | $Chlz_{D1}$ | $P_{D1}$ $P_{D2}$ |
| 1 | 14663 | 56 | 115 | 183 | 225 | **245** | 262 | **330** | 447 |
| 2 | | 14719 | 59 | **127** | 169 | 189 | 206 | 274 | 391 |
| 3 | | | 14778 | 68 | 110 | **130** | 147 | 215 | **332** |
| 4 | | | | 14846 | 42 | 62 | 79 | 147 | 264 |
| 5 | | | | | 14888 | 20 | 37 | 105 | 222 |
| 6 | | | | | | 14908 | 17 | **85** | 202 |
| 7 | | | | | | | 14925 | 68 | 185 |
| 8 | | | | | | | | 14993 | 117 |
| 9 | | | | | | | | | 15110 |

**Table S2:** Exciton energies and difference frequencies (in cm$^{-1}$) in the Novoderezhkin Model (40). Difference frequencies matching the observed coherence frequencies (within 10 cm$^{-1}$) are marked in red.